# Large enhancement of spin-orbit torques under a MHz modulation due to phonon-magnon coupling

Hanying Zhang[1,2], Qianwen Zhao[1,2], Baiqing Jiang[1,2], Yuan Wang[1,2], Tunan Xie[1,2], Kaihua Lou[1,2], ChaoChao Xia[1,3], and C. Bi[1,2,3*]

[1]Laboratory of Microelectronic Devices & Integrated Technology, Institute of Microelectronics, Chinese Academy of Sciences, Beijing 100029, China

[2]University of Chinese Academy of Sciences, Beijing 100049, China

[3]School of Microelectronics, University of Science and Technology of China, Hefei 230026, China

*clab@ime.ac.cn

**Abstract**

The discovery of spin-orbit torques (SOTs) generated through the spin Hall or Rashba effects provides an alternative write approach for magnetic random-access memory (MRAM), igniting the development of spin-orbitronics in recent years. Quantitative characterization of SOTs highly relies on the SOT-driven ferromagnetic resonance (ST-FMR), where a modulated microwave current is used to generate ac SOTs and the modulation-frequency is usually less than 100 kHz (the limit of conventional lock-in amplifiers). Here we have investigated the SOT of typical SOT material/ferromagnet bilayers in an extended modulation-frequency range, up to MHz, by developing the ST-FMR measurement. Remarkably, we found that the measured SOTs are enhanced about three times in the MHz range, which cannot be explained according to present SOT theory. We attribute the enhancement of SOT to additional magnon excitations due to phonon-magnon coupling, which is also reflected in the slight changes of resonant field and linewidth in the acquired ST-FMR spectra, corresponding to the modifications of effective magnetization and damping constant, respectively. Our results indicate that the write current of SOT-MRAM may be reduced with the assistant of phonon-magnon coupling.

Magnetic random-access memory (MRAM) has attracted increasing research attentions from both industry and academia recently, especially, when the current cache memory technology that is based on the static random-access memory (SRAM) consisted of transistors cannot be further scaled beyond the 7 nm technology node[1]. At present, a 32 Mb MRAM with access time less than 5.9 ns has been demonstrated[2], where both the capacity and access time have been approaching the requirements of L3 cache memory. To further increase the write speed and endurance of MRAM, a spin-orbit torque (SOT) driven write scheme has been proposed since there is no incubation time (around 1-2 ns) compared to classic spin-transfer torque (STT) write process[3,4] and the high write current does not need to pass through the MgO tunnel layer[5]. This is because SOT is generated within the adjacent SOT layer by applied write currents and the corresponding spin polarization is originally perpendicular to the magnetization of storage layer, instead of a collinear spin polarization in the STT scheme (only nonlinear spin polarization contributes magnetization switching and thus there is a thermally excited incubation process for STT switching to generate the nonlinear spin polarization)[6]. Practically, there are still many fundamental challenges in SOT-MRAM. In additional to the well-known field-free switching obstacle, high critical switching current density due to low SOT efficiency and the elusive SOT-switching mechanisms also need to be addressed[7–9]. Even for the magnitude of SOT, which dominates the magnetization switching process during write operations and thus determines write efficiencies, contradictory values are given by using different characterization technologies and research groups[10–12]. For instances, the spin Hall angle ($\theta_{SH}$) describing SOT efficiency of a classic SOT material, Pt, the measured $\theta_{SH}$ values are quite inconsistent. Azevedo et al and Liu et al reported $\theta_{SH}$ = 0.08[11] and 0.056[10], respectively, by investigating magnetization dynamics excited under SOTs, while Kimura et al reported $\theta_{SH}$ = 0.0037 by measuring spin transport in a non-local spin valve structure[12].

So far, many efforts have been devoted to the improvement of $\theta_{SH}$ for reducing SOT switching current[13–15]. Generally, SOT is an intrinsic attribute and cannot be modified once the material is formed. Therefore, it is usually improved by doping during material deposition[16–18] or altering the SOT material/ferromagnet (FM) interfaces to reduce spin memory loss[19] and increase spin transparency[20,21]. Moreover, extrinsic effects like magnon splitting[22,23] and magnon-phonon coupling[24–27] have been demonstrated to induce enhanced spin transport from FM to the SOT material, which may also enhance the spin transport in a reverse direction, that is, SOT efficiencies due to spin transport from the SOT material to FM even it has not been discussed in theory and experiment. Here we report the observation of large enhancement of SOT efficiencies, about three times, under a MHz modulation, hinting the role of possible phonon-magnon coupling for enhancing spin transport from the SOT material to FMs.

We have characterized the SOT by using SOT-driven ferromagnetic resonance (ST-FMR), which has been proven to be a sophisticated technique with self-calibration[10] to avoid the disturbance from the Nernst effects or unidirectional magnetoresistance (USMR) in second-harmonic measurements[28–31]. In the ST-FMR measurement, a microwave current passing through the SOT material generates an ac SOT, which induces FMR of the adjacent FM. According to the detected FMR signal as a function of the applied magnetic field (H), which consists of a symmetric ($\Delta H^2/[\Delta H^2 + (H - H_{res})^2]$) and an asymmetric component ($\Delta H(H - H_{res})/[\Delta H^2 + (H - H_{res})^2]$) with the resonant field $H_{res}$ and linewidth $\Delta H$, $\theta_{SH}$ can be calculated through the following equation[10]:

$$\theta_{SH} = \frac{S}{A} \frac{e\mu_0 M_s t d}{\hbar} [1 + (4\pi M_{eff}/H)]^{1/2} \qquad (1)$$

where $S$ and $A$ are the coefficients of the symmetric and asymmetric components, $\mu_0$ is the vacuum permeability, $M_s$ is the saturation magnetization of FM, $\hbar$ is the Planck constant, $4\pi M_{eff}$ is the demagnetization field, and $t$ and $d$ are the thicknesses of FM and SOT layer, respectively. As shown in Fig. 1(a), practical measurements usually adopt a modulated microwave to improve the signal-to-noise ratio and the modulation-frequency ($f_{mod}$) is less than 100 kHz (the limit of conventional lock-in amplifiers). In the $f_{mod} \geq 100$ kHz range where the modulated spin current may interact with phonons or magnons, SOTs have never been investigated due to the lack of effective detection means. By employing a spectrum analyzer, we have detected the modulated SOT signals with $f_{mod}$ up to 2 MHz and investigated the possible phonon contributions to SOTs.

The improved ST-FMR measurement is schematically shown in Fig. 1(b). As demonstrated in the optical image of Fig. 1(b), the ground-signal-ground coplanar waveguide (CPW) was designed with two open ends for injecting and picking up microwave signals, respectively, by using RF probes. The SOT material/FM samples (microstrips with the dimension of 50 μm × 5 μm) are placed in the center of signal lines. The angle between microstrips and H keeps at 45º. The modulated microwave from the signal generator is injected into the CPW through one probe during measurements, and then, the microwave current propagates in the microstrip and finally is picked up by another probe. In contrary to the lock-in measurements shown in Fig. 1(a), all spectrum components of the picked-up microwave signals are directly resolved by using a spectrum analyzer. Typically, there are four peaks at the frequency of $f$, $f/2$, $2f$, and $f_{mod}$ that can be clearly detected by the spectrum analyzer. Here, $f$ is the frequency of input microwave. According to the ST-FMR theory, the detected spectrum components with the frequency of $f$, $f_{mod}$, and $2f$ can reflect magnetization dynamics excited by SOTs, in which the $f$ component mainly corresponds to the microwave absorption, while the $f_{mod}$ and $2f$ signals arise from the resistance oscillation of FM due to magnetization precession. Here we only consider the signals at $f_{mod}$, while $2f$ signals will be discussed in other works. In general, the signal at $f_{mod}$ detected by spectrum analyzers is the same as that detected by lock-in measurements except that the magnitude may be different due to impedance mismatch (lock-in detection requires a bias-tee as illustrated by the dash square in Fig. 1(a) and the input resistance of lock-in amplifier is about 1 MΩ, while the spectrum analyzer is directly connected to RF probes with input resistance of 50 Ω). The comparison of two measurements when $f_{mod}$ < 100 kHz will be performed below. Since the spectrum analyzer can detect signals up to GHz, the developed measurement extends the $f_{mod}$ of ST-FMR in a broader frequency range.

A typical SOT structure, Pt $d_{Pt}$/Py 4 nm (Pt/Py), was used to investigate SOTs at $f_{mod}$ up to MHz, where Pt is the SOT layer, Py is the FM, and 0.5 nm ≤ $d_{Pt}$ ≤ 3 nm is the thickness of the Pt layer. The Pt/Py bilayers as well as a 5 nm $SiO_2$ capping layer were deposited on thermally oxidized Si substrates by using magnetron sputtering. Other SOT structures including Pt 3 nm/CoFeB 10 nm (Pt/CoFeB) and W 8 nm/CoFeB 10 nm (W/CoFeB) were also fabricated to further confirm the enhancement of SOT. The base vacuum before sputtering is better than $8 \times 10^{-9}$ Torr. The deposited Pt or W/FM multilayers were then patterned into 70 μm × 5 μm microstrips by using standard photolithography and ion milling processes. The top CPW lines consisted of Ti 20 nm/Au 80 nm were then deposited through a metal lift-off process. As shown in the optical image of Fig. 1(b), the gap in the center of signal lines of CPW was designed to 50 μm to make sure that the effective area of Pt or W/FM bilayers for microwave propagation is about 50 μm × 5 μm.

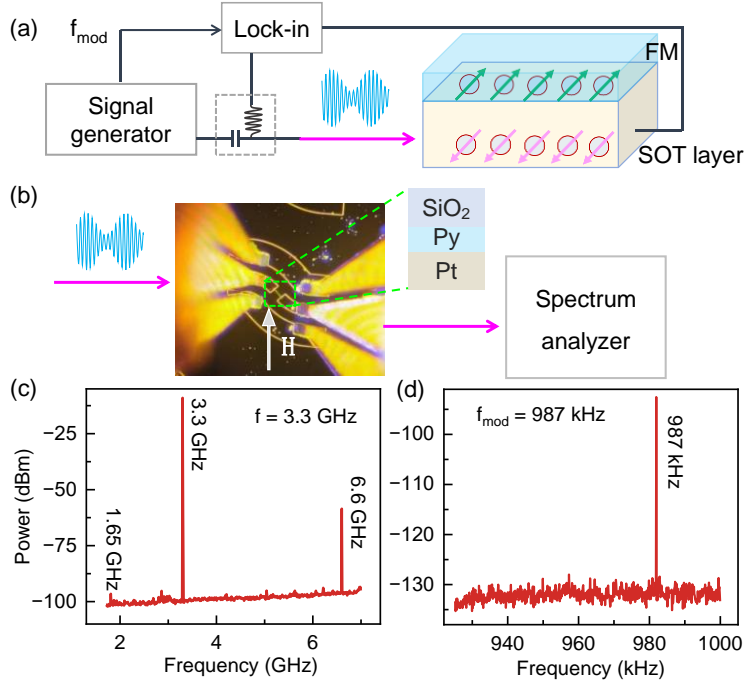

FIG. 1. Experimental configuration. (a) Conventional ST-FMR configuration with a lock-in amplifier. (b) The improved ST-FMR measurements by using a spectrum analyzer. The modulated microwave is injected and picked by using two RF probes as shown in the optical image. (c, d) The acquired spectra of Pt 3/Py 4 nm around (c) 3.3 GHz and (d) 987 kHz by the spectrum analyzer with an $f$ = 3.3 GHz input microwave signal modulated at $f_{mod}$ = 987 kHz.

Figure 1(c) and (d) show the typical spectra of Pt 3/Py 4 nm samples detected by the spectrum analyzer when a modulated microwave signal with $f$ = 3.3 GHz, $f_{mod}$ = 987 kH, and power of 22 dBm was applied, in which four peaks at $f$, $f/2$, $2f$, and $f_{mod}$ can be resolved clearly. The peak amplitude at $f_{mod}$ was then converted to a voltage signal ($V_{f\_mod}$) by considering a 50 Ω input resistance for characterizing SOT according to Eq. (1). To validate the improved ST-FMR measurements, $V_{f\_mod}$ as a function of H measured by using both the lock-in amplifiers and spectrum analyzers is presented in Fig. 2(a) when $f$ = 3.3 GHz and $f_{mod}$ = 67 kHz. As expected, both approaches give the same shape of $V_{f\_mod}$ except the difference on magnitude. For the $V_{f\_mod}$ curve measured by spectrum analyzers, the amplitude is about 5 times smaller than that measured by using lock-in amplifiers. According to Eq. (1), $\theta_{SH}$ is determined by the ratio between symmetric and asymmetric components, $S/A$, which does not relate to the absolute magnitude of $V_{f\_mod}$. The prefect shape overlaps of $V_{f\_mod}$ measured by using two approaches verify the validity of developed approach for characterizing SOTs. Figure 2(b) shows the measured $V_{f\_mod}$ curve under 3.3 GHz input microwave with $f_{mod}$ = 1.33 MHz, in which $f_{mod}$ exceeds the frequency limit of lock-in amplifiers and the $V_{f\_mod}$ curve can still be well deconvoluted into two peaks like that measured by using a lock-in amplifier. According to the ST-FMR theory, the symmetric peak corresponds to damping-like SOT and the asymmetric peak corresponds to field-like SOT[10]. The extracted ΔH and $H_{res}$ under various $f$ are shown in Fig. 2(c) and (d), respectively, in which the calculated damping constant α = 0.018 and $4\pi M_{eff}$ = 0.68 T for Pt 3/Py 4 nm samples by using

$$\Delta H = \frac{2\pi f}{\gamma}\alpha \qquad (2)$$

and

$$f = \frac{\gamma}{2\pi}[H_{res}(H_{res} + 4\pi M_{eff})]^{1/2} \qquad (3)$$

respectively. Here γ is the gyromagnetic ratio. These calculated fundamental attributes approach the previously reported values of Py[10] by a considering a 8% reduction at $f_{mod}$ = 1.33 MHz as demonstrated in Fig. 3(b) and (c).

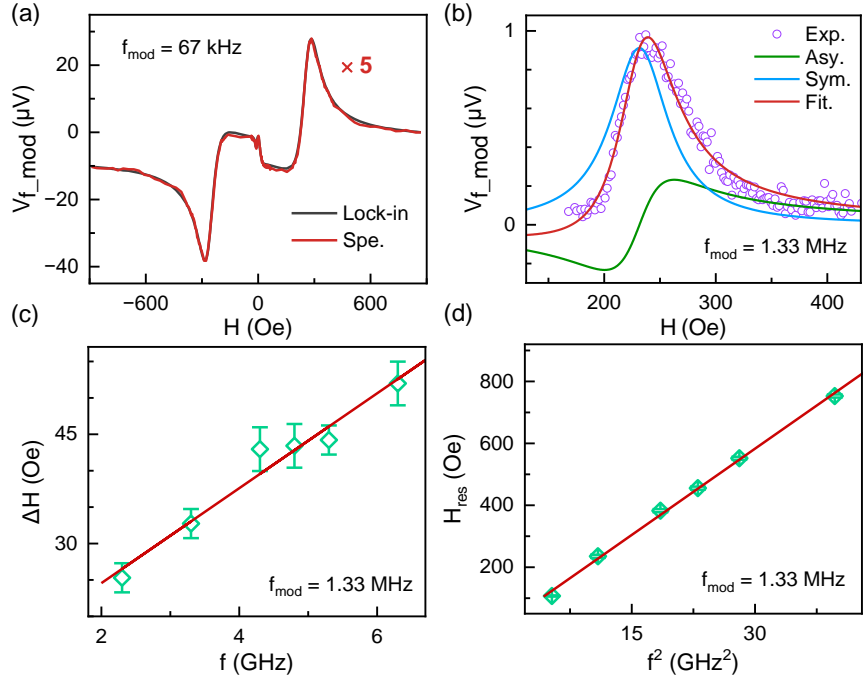

FIG. 2. (a) Comparison of ST-FMR spectra of Pt 3/Py 4 nm measured by spectrum analyzers and lock-in amplifiers. $f$ = 3.3 GHz, $f_{mod}$ = 67 kHz, and power of 15 dBm for input microwave. The red line represents data acquired by spectrum analyzers, which has been amplified 5 times for comparison. (b) Representative ST-FMR spectra of Pt 3/Py 4 nm with $f_{mod} \geq$ 100 kHz, which can also be well fitted by using a symmetric and an asymmetric Lorentzian function like those detected by lock-in amplifiers with $f_{mod}$ < 100 kHz. Circles represent experimental data and the solid lines are fitting results. The input microwave: $f$ = 3.3 GHz, $f_{mod}$ = 1.33 MHz, and power of 22 dBm. (c, d) Extracted (c) $\Delta H$ and (d) $H_{res}$ as a function of $f$ with a fixed $f_{mod}$ = 1.33 MHz for Pt 3/Py 4 nm samples.

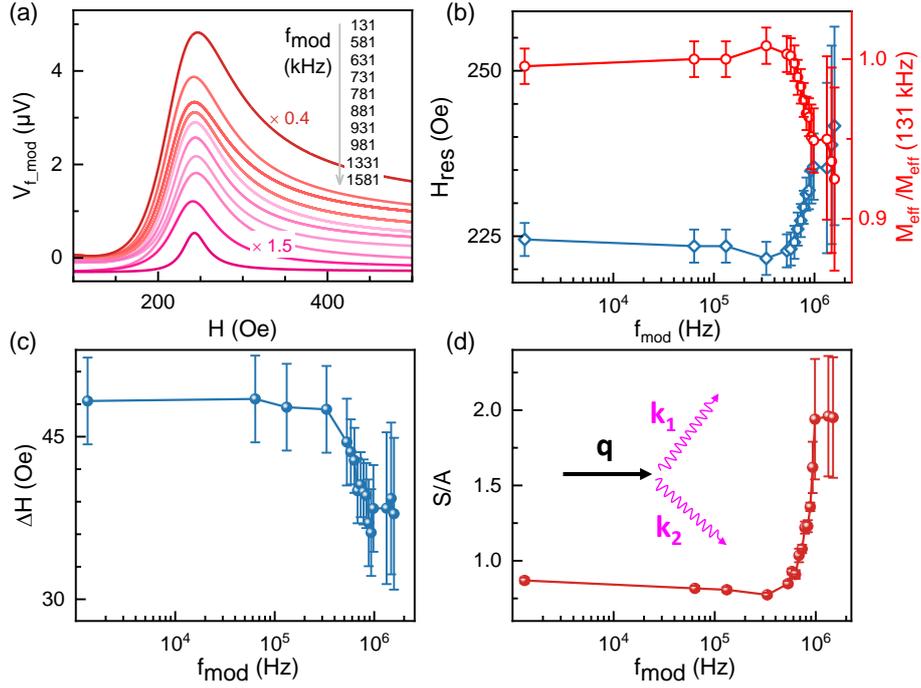

FIG. 3. (a) Evolution of ST-FMR spectra at typical $f_{mod}$ when $f$ = 3.3 GHz for Pt 3/Py 4 nm samples. The power of input microwave keeps at 22 dBm. (b-d) Extracted (b) $H_{res}$, (c) $\Delta H$, and (d) $S/A$ as a function of $f_{mod}$ for Pt 3/Py 4 nm samples ($f$ = 3.3 GHz and power of 22 dBm for input microwave). The change of $M_{eff}$ in the entire $f_{mod}$ range is less than 10% as demonstrated by the red circles in (b). The inset of (d) schematically shows magnon creation through phonon-magnon coupling. **q** and **k** represent the wave vectors of phonon and magnon, respectively.

Remarkable results are shown in Fig. 3(a), where the resonant peak of $V_{f\_mod}$ gradually shifts with increasing $f_{mod}$ and the shape of $V_{f\_mod}$ curves also changes for Pt 3/Py 4 nm samples. To clearly demonstrate the evolution of ST-FMR signals with $f_{mod}$, the $V_{f\_mod}$ curves are deconvoluted into two peaks at each $f_{mod}$ like those in Fig. 2(b). We found that all $V_{f\_mod}$ curves up to $f_{mod}$ = 1.6 MHz can be well fitted by using a symmetric and an asymmetric Lorentzian peak, indicating the validity of this method for SOT characterization up to MHz range. The extracted $H_{res}$, $\Delta H$, and $S/A$ are presented in Fig. 3(b), 3(c), and 3(d), respectively. One can see that all these extracted parameters keep almost constant below $f_{mod}$ = 0.55 MHz and then dramatically change at higher $f_{mod}$. As shown in Fig. 3(b), $H_{res}$ increases from 223 Oe to a saturation value around 242 Oe with increasing $f_{mod}$, indicating a reduced $M_{eff}$ at high $f_{mod}$. According to Eq. (3), the corresponding $M_{eff}$ reduces about 8%, as demonstrated by the calculated relative change of $M_{eff}$ (red circles) in Fig. 3(b). Since the in-plane anisotropic field is much weaker than the demagnetization field in the Pt/Py bilayers, $M_{eff}$ is mainly determined by the saturation magnetization ($M_s$) of Py. The decrease of $M_s$ is usually caused by increasing temperature ($T_a$), generally following the $T_a^{2/3}$ rule in most ferromagnets[32–34]. The $T_a^{2/3}$ rule as well as high order modifications such as $T_a^{7/2}$ and $T_a^4$ terms can be well explained by using the spin-wave theory by considering both magnon-electron and magnon-magnon scattering, in which the decay of spontaneous magnetization is attributed to the magnetization disorder induced by thermally excited magnons with increasing $T_a$[32,34–36]. In our experiments, the input microwave power keeps at 22 dBm for all $f_{mod}$ and the dynamic temperature change because of thermal relaxation occurs in tens of kHz range[37], and therefore, the dramatic increase of temperature around $f_{mod}$ = 0.55 MHz is not expected. Instead, it has been demonstrated that the phonon-magnon coupling can induce additional magnon creation[38–41] when the wave vectors match, $\mathbf{q} = \mathbf{k_1} + \mathbf{k_2}$, as illustrated in the inset of Fig. 3(d). Here, **q** and **k** represent the wave vectors of phonon and magnon, respectively. Experimentally, the phonon-magnon coupling[42–45] and resultant damping change[46,47] and spin transport[24–27] have been reported in many structures. Therefore, the decrease of $M_{eff}$ observed in Fig. 3(b) is more likely due to additional magnon excitation induced by phonon-magnon coupling. As shown in Fig. 3(c), ΔH corresponding to damping constants changes with $f_{mod}$ simultaneously, which is also consistent with magnon excitation[46,47]. However, the detailed mechanism of phonon/magnon creation under a MHz modulation signal through magnetoelastic coupling and quantitative description of resultant magnon contributions on $M_{eff}$ and damping constants require further theoretical calculations.

Remarkably, as shown in Fig. 3(d), the *S/A* ratio that directly determines SOT efficiencies according to Eq. (1) also increases significantly with increasing $f_{mod}$. As mentioned above, it is generally believed that SOT efficiencies are determined by the intrinsic $\theta_{SH}$ of SOT materials[14–18] and the interface transparency of SOT material/FM[19–21], both of which do not change with $f_{mod}$. On the other hand, the inverse effects of SOT, the spin pumping[22,23] and spin Seebeck effects[25–27,48] have demonstrated the crucial role of magnon contribution on spin transport. Therefore, combining the magnon contribution to $M_{eff}$ and ΔH as discussed above in Fig. 3(b) and 3(c), Fig. 3(d) indicates that magnon excitation in the FM also plays a crucial role on the effective SOTs, similar to that in the inverse effects such as spin pumping[22,23] and the spin Seebeck effects[24–26]. In fact, the recently discovered USMR has highlighted the relationship between injected spin current and magnon excitation in the FM even it has not been utilized for evaluating SOTs by considering additional magnon contributions[31,49]. Moreover, recent works show that the SOT efficiencies strongly depend on the composition of FM layers[50–52], which can be caused by the change of magnon dispersion. Figure 4(a) shows the calculated $\theta_{SH}$ as a function of $f_{mod}$ by using Eq. (1), in which the enhancements of SOT at a high $f_{mod}$ are also observed in other SOT bilayers with different $d_{Pt}$, SOT materials, and FMs. When $f_{mod}$ < 0.1 MHz, the calculated $\theta_{SH}$ keeps almost constant for all samples. For Pt 3/Py 4 nm and W 8 nm/CoFeB 10 nm, $\theta_{SH} \approx 0.067$ and -0.28 are close the reported $\theta_{SH}$ values of Pt and W in other works[10,14], respectively, verifying the reliability of developed ST-FMR measurements for qualifying $\theta_{SH}$. With increasing $f_{mod}$, $\theta_{SH}$ increases up to 0.19 and –0.97 for Pt 3/Py 4 nm and W 8 nm/CoFeB 10 nm, respectively, about three times enhancement, when $f_{mod}$ reaches 1.02 MHz. The calculated $\theta_{SH}$ of Pt 3/CoFeB 10 nm about 0.12 when $f_{mod}$ < 0.1 MHz is also consistent with other works by using CoFeB as the FM[53]. The estimated $\theta_{SH}$ varies with different FMs may be caused by interfacial spin transparency[21]. Figure 4(b) shows the calculated $\theta_{SH}$ in Pt/Py bilayers as a function of $d_{Pt}$ at two typical $f_{mod}$ = 1.31 kHz and 1.13 MHz, both of which follow $\theta_{SH} = \theta_{\infty}[1 - \text{sech}(d_{Pt}/\lambda_s)]$ with a saturation value $\theta_{\infty}$ corresponding to the $\theta_{SH}$ when $d_{Pt}$ is much larger than the spin diffusion length $\lambda_s$. When $f_{mod}$ = 1.31 kHz, fitting results give $\theta_{\infty} = 0.077$ and $\lambda_s = 1.17$ nm, similar to previously reported values[54,55]. When $f_{mod}$ = 1.13 MHz, $\theta_{\infty} = 0.24$ and reasonable $\lambda_s = 1.55$ nm are determined, indicating that $\lambda_s$ also depends on $f_{mod}$. It should be noted that the other parameters except $M_s$ and $M_{eff}$ in Eq. (1) do not change and $M_{eff}$ only varies about 10% with increasing $f_{mod}$, and thus, the three times increase of $\theta_{SH}$ cannot be induced by the variation of $M_{eff}$ only in the absence of additional interaction between injected spin current and magnon excitation as discussed above.

Contrary to material or interface engineering for improvement of SOT efficiencies that may cause a large reduction of tunnel magnetoresistance (TMR) in deposited magnetic tunnel junctions (MTJ, core structure of MRAM), our results indicate that the SOT efficiencies can be enhanced by introducing phonon-magnon coupling without modification of the well-developed high-TMR MTJ structures[3,8]. These results may pave an avenue for reducing critical switching current of SOT devices with the assistant of phonon excitation in practical applications. Although the phonon-magnon coupling demonstrated here occurs in the MHz range, it has been shown that the frequency of phonon as well as the phonon-magnon coupling can be extended to the GHz range by using those substrates with fast electrostrictive responses[44–46,56]. In addition, the field-free magnetization switching has been demonstrated in the structures involving strong electrostrictive materials[57–60] in which the inversion symmetry can be broken through strain gradients manipulated by electric fields[58], and thus, energy efficient field-free SOT switching with high write speed can be expected in such structures.

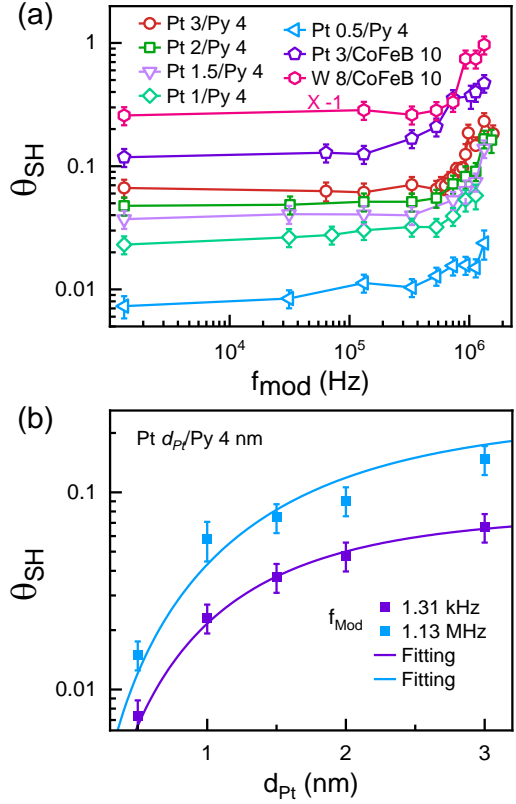

FIG. 4. (a) The calculated $\theta_{SH}$ as a function of $f_{mod}$ in Pt/Py, Pt/CoFeB, and W/CoFeB bilayers with $f$ = 3.3 GHz (Py samples) and 5.3 GHz (CoFeB samples) and power of 22 dBm. (b) The Pt thickness dependence of calculated $\theta_{SH}$ at two typical $f_{mod}$ = 1.31 kHz and 1.13 MHz in Pt $d_{Pt}$/Py 4 nm bilayers. The solid lines are fitting results by using $\theta_{SH} = \theta_{\infty}[1 - \text{sech}(d_{Pt}/\lambda_s)]$.

In conclusion, we have demonstrated that the SOT efficiency can be increased significantly under a MHz modulated microwave excitation, which was directly characterized by using an improved ST-FMR technique. Compared to the 100 kHz limit of conventional ST-FMR measurement by using a lock-in amplifier, the developed ST-FMR promises the detection of SOT with $f_{mod}$ up to MHz or even GHz. When $f_{mod}$ < 0.1 MHz, our results show that θ$_{SH}$ keeps a similar value as previous works, but it dramatically increases about three times when $f_{mod}$ > 1.02 MHz. Correspondingly, both H$_{res}$ and ΔH also show a pronounced change with increasing $f_{mod}$, indicating strong magnon contributions to magnetization dynamics at a higher $f_{mod}$. We attribute the additional magnon excitation to the phonon-magnon coupling. The enhanced SOT efficiency is further explained by considering interactions between spin current and magnon, similar to the mechanisms observed in the USMR, spin pumping, and spin Seebeck effects. On the other hand, the three times enhancement of θ$_{SH}$ is much more efficient than the other material or interface engineering means, which may be used for reducing write current of SOT-MRAM without material or structure modifications in the current device architectures to avoid possible TMR degradation.

**Acknowledgments**

This work is supported by Beijing Natural Science Foundation (Grant No. Z230006), the National Key R&D Program of China (Grant No. 2019YFB2005800 and 2018YFA0701500), the National Natural Science Foundation of China (Grant No. 61974160, 61821091, and 61888102), and the Strategic Priority Research Program of the Chinese Academy of Sciences (Grant No. XDB44000000).

**AUTHOR DECLARATIONS**

**Conflict of Interest**

The authors have no conflicts to disclose.

**DATA AVAILABILITY**

The data that support the findings of this study are available from the corresponding author upon reasonable request.

**References**

[1] P. Weckx, J. Ryckaert, V. Putcha, A. De Keersgieter, J. Boemmels, P. Schuddinck, D. Jang, D. Yakimets, M.G. Bardon, L. Ragnarsson, P. Raghavan, R.R. Kim, A. Spessot, D. Verkest, and A. Mocuta, "Stacked nanosheet fork architecture for SRAM design and device co-optimization toward 3nm," Tech. Dig. - Int. Electron Devices Meet. IEDM, 20.5.1-20.5.4 (2018).

[2] T. Shimoi, K. Matsubara, T. Saito, T. Ogawa, Y. Taito, Y. Kaneda, M. Izuna, K. Takeda, H. Mitani, T. Ito, and T. Kono, in *VLSI* (Institute of Electrical and Electronics Engineers Inc., 2022), pp. 134–135.

[3] K. Garello, F. Yasin, S. Couet, L. Souriau, J. Swerts, S. Rao, S. Van Beek, W. Kim, E. Liu, S. Kundu, D. Tsvetanova, K. Croes, N. Jossart, E. Grimaldi, M. Baumgartner, D. Crotti, A. Fumémont, P. Gambardella, and G.S. Kar, "SOT-MRAM 300MM Integration for Low Power and Ultrafast Embedded Memories," IEEE Symp. VLSI Circuits, Dig. Tech. Pap. **2018-June**, 81–82 (2018).

[4] D.C. Ralph, and M.D. Stiles, "Spin transfer torques," J. Magn. Magn. Mater. **320**(7), 1190–1216 (2008).

[5] L. Liu, C.F. Pai, Y. Li, H.W. Tseng, D.C. Ralph, and R.A. Buhrman, "Spin-torque switching with the giant spin hall effect of tantalum," Science (80-. ). **336**(6081), 555–558 (2012).

[6] K.S. Lee, S.W. Lee, B.C. Min, and K.J. Lee, "Threshold current for switching of a perpendicular magnetic layer induced by spin Hall effect," Appl. Phys. Lett. **102**(11), 112410 (2013).

[7] T. Simsek, "Field-Free Spin-Orbit Torque Switching in Magnetic Tunnel Junction Structures with Stray Fields," IEEE Magn. Lett. **12**, 4500305 (2021).

[8] V. Krizakova, M. Perumkunnil, S. Couet, P. Gambardella, and K. Garello, "Spin-orbit torque switching of magnetic tunnel junctions for memory applications," J. Magn. Magn. Mater. **562**, 169692 (2022).

[9] C.Y. Hu, and C.F. Pai, "Benchmarking of Spin–Orbit Torque Switching Efficiency in Pt Alloys," Adv. Quantum Technol. **3**(8), 2000024 (2020).

[10] L. Liu, T. Moriyama, D.C. Ralph, and R.A. Buhrman, "Spin-Torque Ferromagnetic

Resonance Induced by the Spin Hall Effect," Phys. Rev. Lett. **106**(3), 036601 (2011).

[11] A. Azevedo, L.H. Vilela-Leão, R.L. Rodríguez-Suárez, A.F. Lacerda Santos, and S.M. Rezende, "Spin pumping and anisotropic magnetoresistance voltages in magnetic bilayers: Theory and experiment," Phys. Rev. B **83**(14), 144402 (2011).

[12] T. Kimura, Y. Otani, T. Sato, S. Takahashi, and S. Maekawa, "Room-temperature reversible spin hall effect," Phys. Rev. Lett. **98**(15), 156601 (2007).

[13] L. Zhu, D.C. Ralph, and R.A. Buhrman, "Maximizing spin-orbit torque generated by the spin Hall effect of Pt," APL Mater. **8**(3), (2021).

[14] C.-F. Pai, L. Liu, Y. Li, H.W. Tseng, D.C. Ralph, and R.A. Buhrman, "Spin transfer torque devices utilizing the giant spin Hall effect of tungsten," Appl. Phys. Lett. **101**(12), 122404 (2012).

[15] M. DC, D.-F. Shao, V.D.-H. Hou, A. Vailionis, P. Quarterman, A. Habiboglu, M.B. Venuti, F. Xue, Y.-L. Huang, C.-M. Lee, M. Miura, B. Kirby, C. Bi, X. Li, Y. Deng, S.-J. Lin, W. Tsai, S. Eley, W.-G. Wang, J.A. Borchers, E.Y. Tsymbal, and S.X. Wang, "Observation of anti-damping spin–orbit torques generated by in-plane and out-of-plane spin polarizations in MnPd3," Nat. Mater. **22**(5), 591–598 (2023).

[16] H. Masuda, T. Seki, Y.C. Lau, T. Kubota, and K. Takanashi, "Interlayer exchange coupling and spin Hall effect through an Ir-doped Cu nonmagnetic layer," Phys. Rev. B **101**(22), 224413 (2020).

[17] X. Sui, C. Wang, J. Kim, J. Wang, S.H. Rhim, W. Duan, and N. Kioussis, "Giant enhancement of the intrinsic spin Hall conductivity in β-tungsten via substitutional doping," Phys. Rev. B **96**(24), 241105 (2017).

[18] D. Qu, T. Higo, T. Nishikawa, K. Matsumoto, K. Kondou, D. Nishio-Hamane, R. Ishii, P.K. Muduli, Y. Otani, and S. Nakatsuji, "Large enhancement of the spin Hall effect in Mn metal by Sn doping," Phys. Rev. Mater. **2**(10), 102001 (2018).

[19] J.C. Rojas-Sánchez, N. Reyren, P. Laczkowski, W. Savero, J.P. Attané, C. Deranlot, M. Jamet, J.M. George, L. Vila, and H. Jaffrès, "Spin pumping and inverse spin hall effect in platinum: The essential role of spin-memory loss at metallic interfaces," Phys. Rev. Lett. **112**(10), 106602 (2014).

[20] L. Zhu, L. Zhu, and R.A. Buhrman, "Fully Spin-Transparent Magnetic Interfaces Enabled by the Insertion of a Thin Paramagnetic NiO Layer," Phys. Rev. Lett. **126**(10), 107204 (2021).

[21] W. Zhang, W. Han, X. Jiang, S.H. Yang, and S.S.P. Parkin, "Role of transparency of platinum–ferromagnet interfaces in determining the intrinsic magnitude of the spin Hall effect," Nat. Phys. **11**(6), 496–502 (2015).

[22] H. Kurebayashi, O. Dzyapko, V.E. Demidov, D. Fang, A.J. Ferguson, and S.O. Demokritov, "Controlled enhancement of spin-current emission by three-magnon splitting," Nat. Mater. **10**(9), 660–664 (2011).

[23] H. Sakimura, T. Tashiro, and K. Ando, "Nonlinear spin-current enhancement enabled by spin-damping tuning," Nat. Commun. **5**(1), 5730 (2014).

[24] C.M. Jaworski, J. Yang, S. MacK, D.D. Awschalom, R.C. Myers, and J.P. Heremans, "Spin-seebeck effect: A phonon driven spin distribution," Phys. Rev. Lett. **106**(18), 186601 (2011).

[25] K. Uchida, H. Adachi, T. An, T. Ota, M. Toda, B. Hillebrands, S. Maekawa, and E. Saitoh, "Long-range spin Seebeck effect and acoustic spin pumping," Nat. Mater. **10**(10), 737–741 (2011).

[26] H. Adachi, K.I. Uchida, E. Saitoh, J.I. Ohe, S. Takahashi, and S. Maekawa, "Gigantic enhancement of spin Seebeck effect by phonon drag," Appl. Phys. Lett. **97**(25), 252506 (2010).

[27] K. Uchida, T. Ota, H. Adachi, J. Xiao, T. Nonaka, Y. Kajiwara, G.E.W. Bauer, S. Maekawa, and E. Saitoh, "Thermal spin pumping and magnon-phonon-mediated spin-Seebeck effect," J. Appl. Phys. **111**(10), 103903 (2012).

[28] K. Garello, I.M. Miron, C.O. Avci, F. Freimuth, Y. Mokrousov, S. Blügel, S. Auffret, O. Boulle, G. Gaudin, and P. Gambardella, "Symmetry and magnitude of spin–orbit torques in ferromagnetic heterostructures," Nat. Nanotechnol. **8**(8), 587–593 (2013).

[29] J. Kim, J. Sinha, M. Hayashi, M. Yamanouchi, S. Fukami, T. Suzuki, S. Mitani, and H. Ohno, "Layer thickness dependence of the current-induced effective field vector in Ta|CoFeB|MgO," Nat. Mater. **12**(3), 240–245 (2013).

[30] K. Lou, Q. Zhao, B. Jiang, and C. Bi, "Large Anomalous Unidirectional Magnetoresistance in a Single Ferromagnetic Layer," Phys. Rev. Appl. **17**(6), 064052 (2022).

[31] C.O. Avci, J. Mendil, G.S.D. Beach, and P. Gambardella, "Origins of the Unidirectional Spin

Hall Magnetoresistance in Metallic Bilayers," Phys. Rev. Lett. **121**(8), 087207 (2018).

[32] A.T. Aldred, "Temperature dependence of the magnetization of nickel," Phys. Rev. B **11**(7), 2597 (1975).

[33] K. Hüller, "The spin wave excitations and the temperature dependence of the magnetization in iron, cobalt, nickel and their alloys," J. Magn. Magn. Mater. **61**(3), 347–358 (1986).

[34] H.A. Mook, J.W. Lynn, and R.M. Nicklow, "Magnetic Excitations in Nickel and Iron," AIP Conf. Proc. **18**(1), 781–793 (1974).

[35] B.E. Argyle, S.H. Charap, E.W. Pugh, B.E. Argyle, S.H. Charap, and E.W. Pugh, "Deviations from T32 Law for Magnetization of Ferrometals: Ni, Fe, and Fe+3% Si," Phys. Rev. **132**(5), 2051–2062 (1963).

[36] M.D. Kuzmin, "Shape of temperature dependence of spontaneous magnetization of ferromagnets: Quantitative analysis," Phys. Rev. Lett. **94**(10), 107204 (2005).

[37] Y.S. Gui, N. Mecking, A. Wirthmann, L.H. Bai, and C.M. Hu, "Electrical detection of the ferromagnetic resonance: Spin-rectification versus bolometric effect," Appl. Phys. Lett. **91**(8), 082503 (2007).

[38] S. Streib, N. Vidal-Silva, K. Shen, and G.E.W. Bauer, "Magnon-phonon interactions in magnetic insulators," Phys. Rev. B **99**(18), 184442 (2019).

[39] K. An, A.N. Litvinenko, R. Kohno, A.A. Fuad, V. V. Naletov, L. Vila, U. Ebels, G. De Loubens, H. Hurdequint, N. Beaulieu, J. Ben Youssef, N. Vukadinovic, G.E.W. Bauer, A.N. Slavin, V.S. Tiberkevich, and O. Klein, "Coherent long-range transfer of angular momentum between magnon Kittel modes by phonons," Phys. Rev. B **101**(6), 060407 (2020).

[40] C. Kittel, "Interaction of Spin Waves and Ultrasonic Waves in Ferromagnetic Crystals," Phys. Rev. **110**(4), 836 (1958).

[41] C. Zhao, Z. Zhang, Y. Li, W. Zhang, J.E. Pearson, R. Divan, Q. Liu, V. Novosad, J. Wang, and A. Hoffmann, "Direct Imaging of Resonant Phonon-Magnon Coupling," Phys. Rev. Appl. **15**(1), 014052 (2021).

[42] R. Weber, "Magnon-Phonon Coupling in Metallic Films," Phys. Rev. **169**(2), 451 (1968).

[43] C. Berk, M. Jaris, W. Yang, S. Dhuey, S. Cabrini, and H. Schmidt, "Strongly coupled magnon–phonon dynamics in a single nanomagnet," Nat. Commun. **10**(1), 2652 (2019).

[44] J. Holanda, D.S. Maior, A. Azevedo, and S.M. Rezende, "Detecting the phonon spin in magnon–phonon conversion experiments," Nat. Phys. **14**(5), 500–506 (2018).

[45] B. Heinrich, and J.F. Cochran, "Phonon assisted magnon transmission through nickel at 24 GHz," J. Appl. Phys. **50**(B3), 2440–2442 (1979).

[46] R. Schlitz, L. Siegl, T. Sato, W. Yu, G.E.W. Bauer, H. Huebl, and S.T.B. Goennenwein, "Magnetization dynamics affected by phonon pumping," Phys. Rev. B **106**(1), 014407 (2022).

[47] S. Streib, H. Keshtgar, and G.E.W. Bauer, "Damping of Magnetization Dynamics by Phonon Pumping," Phys. Rev. Lett. **121**(2), 027202 (2018).

[48] J. Xiao, G.E.W. Bauer, K.C. Uchida, E. Saitoh, and S. Maekawa, "Theory of magnon-driven spin Seebeck effect," Phys. Rev. B **81**(21), 214418 (2010).

[49] K. Yasuda, A. Tsukazaki, R. Yoshimi, K.S. Takahashi, M. Kawasaki, and Y. Tokura, "Large Unidirectional Magnetoresistance in a Magnetic Topological Insulator," Phys. Rev. Lett. **117**(12), 127202 (2016).

[50] L. Zhu, and D.C. Ralph, "Strong variation of spin-orbit torques with relative spin relaxation rates in ferrimagnets," Nat. Commun. **14**(1), 1778 (2023).

[51] J. Finley, and L. Liu, "Spin-Orbit-Torque Efficiency in Compensated Ferrimagnetic Cobalt-Terbium Alloys," Phys. Rev. Appl. **6**(5), 054001 (2016).

[52] R. Mishra, J. Yu, X. Qiu, M. Motapothula, T. Venkatesan, and H. Yang, "Anomalous Current-Induced Spin Torques in Ferrimagnets near Compensation," Phys. Rev. Lett. **118**, 167201 (2017).

[53] Z. Xu, G.D.H. Wong, J. Tang, E. Liu, W. Gan, F. Xu, and W.S. Lew, "Large spin Hall angle enhanced by nitrogen incorporation in Pt films," Appl. Phys. Lett. **118**(6), 062406 (2021).

[54] K. Kondou, H. Sukegawa, S. Mitani, K. Tsukagoshi, and S. Kasai, "Evaluation of spin Hall angle and spin diffusion length by using spin current-induced ferromagnetic resonance," Appl. Phys. Express **5**(7), 073002 (2012).

[55] Y. Wang, P. Deorani, X. Qiu, J.H. Kwon, and H. Yang, "Determination of intrinsic spin Hall angle in Pt," Appl. Phys. Lett. **105**(15), 152412 (2014).

[56] T. Vasileiadis, J.S. Reparaz, and B. Graczykowski, "Phonon transport in the gigahertz to terahertz range: Confinement, topology, and second sound," J. Appl. Phys. **131**(18), 180901

(2022).

[57] I.S. Camara, J.Y. Duquesne, A. Lemaître, C. Gourdon, and L. Thevenard, "Field-Free Magnetization Switching by an Acoustic Wave," Phys. Rev. Appl. **11**(1), 014045 (2019).

[58] Q. Wang, J. Domann, G. Yu, A. Barra, K.L. Wang, and G.P. Carman, "Strain-Mediated Spin-Orbit-Torque Switching for Magnetic Memory," Phys. Rev. Appl. **10**(3), 034052 (2018).

[59] K. Cai, M. Yang, H. Ju, S. Wang, Y. Ji, B. Li, K.W. Edmonds, Y. Sheng, B. Zhang, N. Zhang, S. Liu, H. Zheng, and K. Wang, "Electric field control of deterministic current-induced magnetization switching in a hybrid ferromagnetic/ferroelectric structure," Nat. Mater. **16**(7), 712–716 (2017).

[60] J.M. Hu, Z. Li, J. Wang, and C.W. Nan, "Electric-field control of strain-mediated magnetoelectric random access memory," J. Appl. Phys. **107**(9), 093912 (2010).